\documentclass[11pt]{article}
\pdfoutput=1
\usepackage{jheppubnew}
\usepackage{amsmath}
\usepackage{slashed}
\usepackage{pstool}
\usepackage{mathtools}
\usepackage{color}
\usepackage{float}
\usepackage{array}
\usepackage{amssymb}
\usepackage{amsthm}
\usepackage{tikz}
\usetikzlibrary{positioning}
\usetikzlibrary{intersections}
\usetikzlibrary{fadings} 

\usetikzlibrary{decorations.pathmorphing}
\usetikzlibrary{decorations.pathreplacing,decorations.markings}
\usepackage{graphicx}
\usepackage{caption}
\usepackage[labelsep=quad]{subcaption}
\usepackage{epstopdf}
	
\usepackage{epsfig}
\usepackage[toc,page]{appendix}
\usepackage{empheq}

\newcommand{\be}{\begin{equation}}
\newcommand{\bea}{\begin{eqnarray}}
\newcommand{\eea}{\end{eqnarray}}
\newcommand{\ba}{\begin{array}}
\newcommand{\ea}{\end{array}}
\newcommand{\ee}{\end{equation}}

\title{Complexity as a holographic probe of strong cosmic censorship}
\author[a]{Mohsen Alishahiha,}
\emailAdd{alishah@ipm.ir}
\affiliation[a]{School of Physics, Institute for Research in Fundamental Sciences (IPM),\\
	P.O. Box 19395-5531, Tehran, Iran\\} 
\author[b]{Souvik Banerjee,}
\emailAdd{souvik.banerjee@physik.uni-wuerzburg.de}
\affiliation[b]{Institut f{\"u}r Theoretische Physik und Astrophysik,
	Julius-Maximilians-Universit{\"a}t W{\"u}rzburg,\\ Am Hubland, 97074 W{\"u}rzburg, Germany\\} 
\author[c,d]{Joshua Kames-King}
\emailAdd{jvakk@yahoo.com}
\affiliation[c]{Bethe Center for Theoretical Physics 
and
Physikalisches Institut der Universitaet Bonn, Nussallee 12, 53115 Bonn, Germany\\} 
\affiliation[d]{Kavli Institute for Theoretical Physics,
	University of California,
	Santa Barbara, California 93106, USA} 
\author[b]{and Emma Loos}
\emailAdd{emma.loos@physik.uni-wuerzburg.de}

\begin{document}

\abstract{ Based on reasonable assumptions, we propose a new expression for Lloyd's 
bound, which confines the complexity growth of charged black holes. We then revisit holographic complexity for 
charged black branes in the presence of a finite cutoff. Using the proposed Lloyd's bound we find a 
relation between the UV and the behind the horizon cutoff. This is found to be consistent with the factorization of the 
partition function at leading order in large N. We argue that the result may be 
thought of as a holographic realization of strong cosmic censorship.
}

\maketitle

\newcommand{\RNum}[1]{\uppercase\expandafter{\romannumeral #1\relax}}
\section{INTRODUCTION}
\label{sec:intro}

In the past decade, there have been a lot of interesting developments in understanding and resolving puzzles related to the interior of black holes, the mysterious part of the spacetime hiding behind the black hole event horizon. In particular, in the context  of the AdS/CFT correspondence, in which  these paradoxes can be given a sharp form in terms of information processing of the boundary conformal field theory (CFT) \cite{Mathur:2009hf, Almheiri:2012rt, Almheiri:2013hfa, Marolf:2013dba}, a lot of effort has been devoted towards resolving the aforementioned problems using the entanglement structure of spacetime \cite{Papadodimas:2012aq, Papadodimas:2013jku}. Some of the resolutions are even instrumental in understanding the unique nature of entanglement in generic systems of quantum gravity \cite{Banerjee:2016mhh, Chowdhury:2020hse}.

These developments motivated a rigorous search for probes both sensitive to the interior of a black hole and which also systematically relate to the evolution of operators in the boundary CFT. Holographic complexity turned out to be one such probe. It was originally proposed in terms of an entangled pair of black holes \cite{Susskind:2014rva}. The pair exchanges information through a virtual wormhole structure, namely, {\it the Einstein-Rosen (ER) bridge} \cite{Maldacena:2013xja}. The bridge growing in time is identified as the holographic complexity growth in this set up. In an anti-de Sitter (AdS) black hole spacetime, such a notion naturally relates the radial depth in the bulk spacetime to the growth of boundary operators. As a consequence, the black hole spacetime can be thought of as an onion shell-like structure with each radial slice corresponding to a particular complexity \cite{Susskind:2014rva} of the dual boundary CFT.
One efficient way to compute the complexity of a holographic state was proposed in \cite{Brown:2015bva, Brown:2015lvg} the ``complexity = action'' (CA) conjecture. In this conjecture, the holographic complexity is given by the on-shell action on the Wheeler-DeWitt (WdW) patch which is the domain of dependence of any Cauchy surface in the bulk which intersects the asymptotic boundary on the time slice, $\Sigma$,
\begin{equation}
\label{eq:CA}
{\cal C}\left(\Sigma\right) = \frac{{\cal I}_{\rm{WdW}}}{\pi \hbar}.
\end{equation}

One interesting tool to investigate the precise relation between complexity and radial depth in a black hole spacetime is the recently proposed duality between AdS spacetimes cut off at a finite radial distance and dual CFTs deformed by an irrelevant operator, known as the {\it $T \bar T$ deformation} \cite{McGough:2016lol, Taylor:2018xcy, Hartman:2018tkw}. $T \bar T$ is a certain
 quadratic combination of the stress-energy tensor of the boundary field theory \cite{Zamolodchikov:2004ce, Smirnov:2016lqw, Cavaglia:2016oda}.
This correspondence is very nontrivially supported by the matching of the energy spectrum measured by an observer at a finite distance away from the black hole in AdS spacetime and that of a $T \bar T$ deformed CFT.\footnote{We note, however, that  one should be  careful 
once we are dealing with a gravity theory with a finite radial cutoff \cite{Witten:2018lgb}. 
It has been shown that the ${\rm T{\bar T}}$ deformation  might be better 
described by  imposing mixed boundary conditions at the asymptotic boundary \cite{Guica:2019nzm}. }


An attempt to explore  the time evolution of holographic complexity for a black hole in AdS with a radial cutoff was made in \cite{Akhavan:2018wla}. It was shown  that in order for this complexity to grow linearly with time with the coefficient approaching a constant value equal to twice the energy of the state, (this is known as the ``Lloyd's bound'' in the literature \cite{Lloyd_2000}
\footnote{In the context of holographic complexity, it is known that Lloyd's bound may actually
be violated \cite{Carmi:2017jqz}. Nonetheless, the violation of Lloyd's bound just modifies the relation between the cutoff
behind the horizon and the UV cutoff at intermediate times, depending on whether at  late times the bound is saturated from above or below. This does not, however, affect the main conclusion of our paper. Because in either case, the saturation is guaranteed at late times and the late-time behavior of complexity growth is controlled by boundary physical charges whose values are affected by a finite UV cutoff.
}) it is necessary to invoke a cutoff behind the horizon as well, with a value fixed by the boundary UV cutoff. The  precise relation between the boundary cutoff and the cutoff behind the horizon
has also been obtained in \cite{Akhavan:2018wla}. The
corresponding relations between the two cutoffs for charged black holes and near extremal charged black branes in AdS were derived in \cite{Hashemi:2019xeq} and in \cite{Alishahiha:2019cib}, respectively. 

Intuitively the relation between the two cutoffs may be understood as follows. In the
context of the CA  proposal, the late-time behavior of complexity growth is 
 entirely given by the on-shell action evaluated on the intersection of the WdW patch with 
 the future interior \cite{Alishahiha:2018lfv}, leading  to an observation that the late-time 
 behavior of holographic complexity is insensitive to the UV cutoff \cite{Akhavan:2018wla}.  
 On the other hand  according to Lloyd's bound \cite{Lloyd_2000}, the late-time behavior
of complexity growth is given in terms of  the energy of the system that is sensitive to 
the finite UV cutoff.  Therefore,   while the physical charges are sensitive to  a UV 
cutoff, the late-time behavior of holographic complexity seems blind to the UV cutoff.  A remedy to resolve this puzzle is to assume that the  UV cutoff will induce a cutoff behind the horizon with a value fixed by the UV cutoff.

 By relating the partition functions inside and outside of the horizon of an eternal black hole using the  ``mirror operator'' construction of Papadodimas and Raju \cite{Papadodimas:2012aq, Papadodimas:2013jku}, the authors of \cite{Akhavan:2019vtt} established that a cutoff at a finite radial distance does indeed imply a cutoff behind the horizon. This guarantees a bulk effective field theory at leading order in the ${1\over N}$ expansion. Remarkably, the relation between the cutoffs obtained this way, exactly matches the one derived in \cite{Akhavan:2018wla}, hinting at a deep connection between radial distance and complexity as well as with the black hole information paradox.

In this paper we shall consider a charged, eternal black brane in AdS. It has an inner horizon in addition to its outer event horizon which makes the causal structure of such spacetime geometries even more rich and interesting. In particular, there has been a long-standing debate regarding the fate of an infalling observer after crossing the event horizon of such black holes. Whether the observer can also cross the inner horizon smoothly is a tricky question since this horizon, being a Cauchy horizon, does not guarantee a unique evolution of smooth initial data. This problem is resolved in classical gravity using the conjecture of ``strong cosmic censorship'' that predicts the eventual collapse of the inner horizon as soon as the infalling observer reaches it. This instability is an artefact of an infinite blue shift effect. It is, however, very difficult to prove this in general particularly beyond the regime of classical gravity. In recent work, \cite{Papadodimas:2019msp}, a quantum test in form of the behavior of boundary correlators was proposed in order to diagnose the smoothness of the inner horizon for charged AdS black holes.

In our study we will set the interior cutoff behind the event horizon but outside the inner horizon and derive a relation between the two cutoffs in two different ways : first by making use of complexity growth and secondly by using the validity of low-energy effective field theory and the factorization of the corresponding Hilbert space.
We will then give a dual interpretation of our result in terms of the emergence of strong cosmic censorship. 

On our way towards deriving the relation between the cutoffs, we will also address a long-standing issue regarding the bound on the late-time growth of complexity.  In the existing literature, there is no unique consensus on the generalization of Lloyd's bound for a charged system. The reason for this apparent ambiguity is that, unlike the uncharged case, in the case of a charged black hole, this bound is hard to ``derive'' from first principles. There have been two proposals based on ``natural expectations'' \cite{Brown:2015lvg, Cai:2016xho} : however, both of them suffer from certain pathologies. The cutoff geometry makes this problem even more complicated. However, in our case since we compute the relation between the cutoffs in two different ways, it can be used as a very nice diagnostic of the correct Lloyd's bound for a charged system. In fact, we will propose a new bound for charged black holes (branes) which apart from being consistent with several limits, namely, the zero cutoff and zero charge limits, is also free from the aforementioned issues associated with previous proposals.

The rest of the paper is organized as follows. In Sec. \ref{sec:complexity} we present the computation of the late-time growth of complexity in detail. Section \ref{sec:Lloyd} will be devoted to the discussion of the generalization of Lloyd's bound to charged black branes. Here we will propose a new bound and compare it with the previously existing bounds in the literature. At the end of that section we will present the relation between the two cutoffs implied by the proposed bound. In Sec. \ref{sec:factorization} we will derive the relation between the cutoffs from a different perspective, namely from the factorization of the partition function. We will show that, provided we use the proposed Lloyd's bound, the relation between the cutoffs in both the approaches match exactly. Section \ref{sec:SCC} is reserved for the interpretation of our results, particularly, in terms of an emergent strong cosmic censorship. We will conclude in Sec. \ref{sec:discussion-outlook} with some comments on the choice of ensembles and also some future outlooks.


\section{COMPLEXITY OF A CHARGED BLACK BRANE WITH CUTOFF}
\label{sec:complexity}
In this section we shall study the complexity growth of an eternal charged black brane solution 
with a finite radial cutoff.\footnote{ We note that the complexity of charged black holes 
in the presence of a finite cutoff has been 
already studied in \cite{Hashemi:2019xeq} (see also \cite{Razaghian:2020bfk}) and 
indeed most parts of this section are a review.  Our aim is to present the results in a new, inspiring form.} To proceed, let us first fix our notation. We will consider the  Einstein-Hilbert-Maxwell bulk action 
\begin{align}
\label{eq:bulkMax} 
S_{\mathrm{bulk}} = 
\frac{1}{16 \pi G_N} \int d^{d+2} x \sqrt{-g} (R - 2 \Lambda-\frac{1}{2}F_{\mu \nu} F^{\mu \nu}) \,,
\end{align}
for which
eternal charged black brane solutions,
for  $d>2$ may be given as follows\footnote{In what follows we only 
consider electric brane solutions in which  the only nonzero components of the electromagnetic 
field strengths are $F^{rt}$ and $F^{tr}$. We have also assumed the radial gauge, $A_r = 0$. For a 
complete discussion on these solutions and how to obtain them, see \cite{Chamblin:1999tk}.} 
\begin{align}
ds^2 = \frac{L^2}{r^2} \bigg( -f(r) dt^2 + \frac{dr^2}{f(r)} + \sum_{i=1}^{d} dx_i^2 
 \bigg)\,, \;\;\text{ with } f(r) = 1 - m r^{d+1} +Q^2 r^{2d}
  \label{eq:RNmetric}
\end{align}
and 
\begin{equation}
  \label{eq:RNgauge}
A_t = \sqrt{\frac{d}{d-1}} Q L \left(r_+^{d-1} - r^{d-1}\right).
\end{equation}
Here $m$ and $Q$ 
are related to the mass and the charge of the black brane, respectively. In particular, the total ADM charge of the system is given by 
\be
\label{ADMQ}
{\cal Q} = \oint *F = \sqrt{d(d-1)} \, \frac{V_d L^{d-1}}{8 \pi G_N} Q,
\ee
where the $d$-form field, $*F$, is the Hodge dual of the electromagnetic field strength tensor 
$F_{\mu\nu} = \partial_\mu A_{\nu}-\partial_\nu A_{\mu}$. $V_d$ denotes the transverse $d$-dimensional volume. $G_N$ is the $(d+2)$-dimensional Newton's constant and $L$ is the AdS 
length scale.

\begin{figure}
\centering
\begin{tikzpicture}
[scale=0.6] 
	\node (I)    at ( 4,0)   {};
\node (II)   at (-4,0)   {};
\node (IIIa)  at (0, 2.5) {};
\node (IV)   at (0,-2.5) {};
\node (IIIb) at (0, 5) {};
\node (V) at (-4,7.9) {};
\node (VI) at (4,7.9) {};
\coordinate(s) at  (0,-1.8);
 \coordinate (sr) at  (25:3.87cm);
 \coordinate (sl) at (-25:-3.87cm) ;
 \coordinate (sur) at  (1.45,3.5);
 \coordinate (sul) at (-1.45,3.5);

 \draw (s)--(sr)node[midway,sloped,below] {$B_2$};
  \draw (s)--(sl)node[midway,sloped,below] {$B_3$};
  \draw (sr)--(sur);
  \draw (sl)--(sul);
  \path[fill=white!85!green] (s) -- (sr)--(sur)--(sul)--(sl)-- cycle;
  \coordinate(si) at  (0,0.);
 \coordinate (sri) at  (45:3.55cm);
 \coordinate (sli) at (-45:-3.55cm) ;
 \coordinate (suri) at  (1.45,3.5);
 \coordinate (suli) at (-1.45,3.5);

 \draw (si)--(sri) ;
  \draw (si)--(sli);
  \draw (sri)--(suri) node[midway,sloped,above] {$B_1$};
  \draw (sli)--(suli) node[midway,sloped,above] {$B_4$};
  \path[fill=white!40!green] (si) -- (sri)--(suri)--(suli)--(sli)-- cycle;
  
\path  
  (VI) +(90:4)  coordinate[]  (VItop)
       +(-90:4) coordinate[] (VIbot)
       +(0:-4)   coordinate                  (VIleft);
       \draw 
      (VItop) --
          node[midway, below, sloped] {}
      (VIleft) ;
      \draw (VIleft) -- 
           node[midway, below, sloped] {}
      (VIbot);
\draw    (VIbot) [decorate,decoration=snake]    --
          node[midway, above, sloped] {}
          node[midway, below left]    {}    
      (VItop) ;
\path  
  (V) +(90:4)  coordinate[]  (Vtop)
       +(-90:4) coordinate[] (Vbot)
       +(0:4)   coordinate                  (Vright);
       \draw 
      (Vtop) --
          node[midway, below, sloped] {}
      (Vright) -- 
          node[midway, below, sloped] {}
      (Vbot) ;
\draw   (Vbot)  [decorate,decoration=snake]    --
          node[midway, above, sloped] {}
          node[midway, below left]    {}    
      (Vtop) ;

\path  
  (II) +(90:4)  coordinate[]  (IItop)
       +(-90:4) coordinate[] (IIbot)
       +(0:4)   coordinate                  (IIright);
       \draw 
      (IItop) --
          node[midway, below, sloped] {}
      (IIright) -- 
          node[midway, below, sloped] {}
      (IIbot) --
          node[midway, above, sloped] {}
          node[midway, below left]    {}    
      (IItop) -- cycle;

\path 
   (I) +(90:4)  coordinate (Itop)
       +(-90:4) coordinate (Ibot)
       +(180:4) coordinate (Ileft)
       ;
\draw  (Ileft)-- (Itop)-- (Ibot) -- (Ileft) -- cycle;


      \draw[blue,decorate, decoration={segment length=1cm, pre=lineto, pre length=0.6cm, post = lineto, post length=0.1cm}] (IItop)  to[bend right=13]  (Itop);
      \node[blue] at (0,3.9) {$r_0$};
      \draw[red,decorate, decoration={segment length=1cm, pre=lineto, pre length=0.6cm, post = lineto, post length=0.1cm}] (Itop)  to[bend right=15]  (Ibot);
      \node at (2.9,0)[red] {$r_c$};
       \node at (-2.9,0)[red] {$r_c$};
 	
 	 \draw[red,decorate, decoration={segment length=1cm, pre=lineto, pre length=0.6cm, post = lineto, post length=0.1cm}] (IItop)  to[bend left=15]  (IIbot);
 	 \node at (-1.8,0)   {};
 	 \node at (1.8,0)   {};
 	 \node at (1.8,8)   {};
 	 \node at (-1.8,8)   {};
 	 \node at (-0.59,0){$J_2$};
 	 \node at (-2.6,2){$J_3$};
 	 \node at (2.6,2) {$J_1$};
 	 \filldraw[black] (0,0) circle (2pt)  {};
 	  \filldraw[black] (-2.5,2.5) circle (2pt)  {};
 	  \filldraw[black] (2.5,2.5) circle (2pt)  {};
 	   \filldraw[black] (1.4,3.55) circle (2pt)  {};
 	  \filldraw[black] (-1.4,3.55) circle (2pt)  {};
 	  \node at (0,-2) {$r_m$};
 	  \draw [white!20!black,dashed](0,3.5) --(0,-1.8) ;
 	  \node [white!20!black] at (0.3,-0.9) {$c$};
 	  \node [white!20!black] at (1.73,0.88) {$b$};
 	  \node [white!20!black] at (0.5,1.88) {$a$};
	  
 	 \node at (5.5,0) {$r=0$};
 	 \node at (-5.5,0) {$r=0$};
 	  \node at (5.5,8) {$r\rightarrow \infty$};
 	 \node at (-5.5,8) {$r\rightarrow \infty$};
 	 \node at (-2.2,-3) {$r_+$};
 	 \node at (2.2,-3) {$r_+$};
 	 \node at (2.2,9.3) {$r_-$};
 	 \node at (-2.2,9.3) {$r_-$};

\end{tikzpicture}
 \caption{The Penrose diagram of the eternal charged black brane with the WdW patch depicted in light green. The intersection of the WdW with the future interior is denoted in a darker green. We have also labeled the four lightlike boundaries by $B_1,...,B_4$ and denoted joint points of these boundaries with each other and with the spacelike surface $r_0$ by black dots. The three lightlike joint points are marked as $J_1, J_2, J_3$. Moreover, we also introduce the three regions $a,b,c$, which break the WdW patch into simple contributions.}
 \label{fig:latetimecompgrowth}
\end{figure}
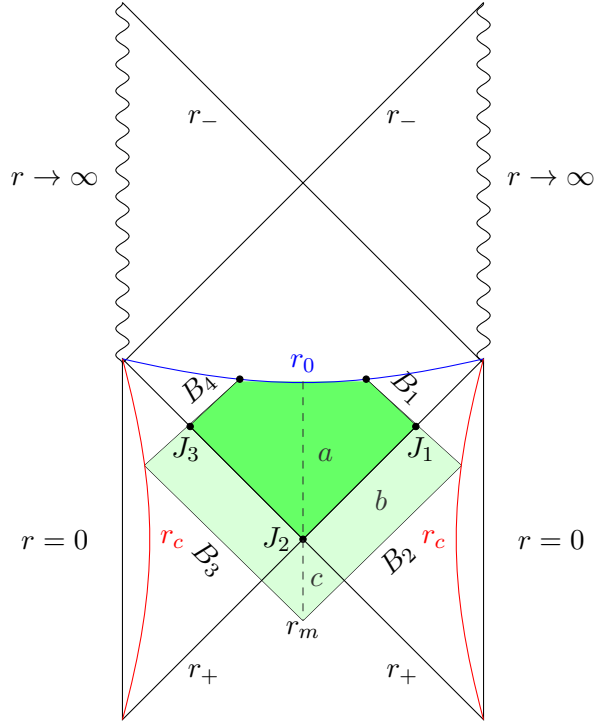

This geometry has two horizons, the outer horizon $r_+$, and the inner horizon $r_-$, as depicted 
in the Penrose diagram in Fig. \ref{fig:latetimecompgrowth}. 
They correspond to the positive real roots of 
the equation $f(r) = 0$, where $f(r)$ is the blackening factor given in \eqref{eq:RNmetric}. In our 
choice of coordinates in which the AdS boundary is located at $r = 0$,
one has $r_+ < r_-$.

 Furthermore, following \cite{Akhavan:2018wla}, we introduce two cutoffs, the UV cutoff at
  $r = r_c$ and a cutoff at $r = r_0$ lying 
 behind the outer horizon.  We place our cutoff, $r_0$ in between the inner and the outer horizons,  
 $r_+<r_0<r_-$. However, in principle, $r_0$ could also lie behind the inner horizon, $r_0>r_-$.  
 For the moment, this is only a choice, but we will elaborate upon and justify this in Sec. \ref{sec:SCC}.

Now we proceed to compute the late-time growth of complexity in the charged black brane 
geometry with the aforementioned two cutoffs. 
We will use the CA proposal given in \eqref{eq:CA},\footnote{Although the procedure of computing complexity using the CA proposal is, by now, quite standard, in order 
to be self-contained, we will present the computations in a rather detailed manner.} which requires the evaluation 
of the on-shell action on the WdW patch associated with a boundary state at a time, $\tau = t_L + 
t_R$. Here $t_L$ is the time at the left boundary and $t_R$ at the right boundary. The full WdW 
patch is the union of the dark and the light green regions in Fig.
\ref{fig:latetimecompgrowth}.\footnote{We are ultimately interested in the late-time 
behavior which means that we effectively restrict to the intersection of the WdW patch with the future interior. This region is highlighted as the  dark 
green patch in Fig. \ref{fig:latetimecompgrowth}.}
  As for the computations in this section, we will closely follow \cite{Akhavan:2018wla} ( see also \cite{Hashemi:2019xeq} and\cite{Alishahiha:2019cib}).

In general, the action on the WdW patch contains the following pieces \cite{Parattu_2016_1, Parattu_2016_2, Lehner_2016}
\begin{align}
\label{IWdW}
{\cal I}_{\rm{WdW}} = I_{\rm{bulk}} + I_{\rm{GH}} + I_{\rm{CT}} + I_{\rm j}\,.
\end{align}
The individual terms on the right-hand side  correspond to the on-shell bulk action, the Gibbons-Hawking surface terms, the counterterms and the contributions coming from the joint points on the WdW patch, respectively. While the bulk contribution can be straightforwardly obtained by evaluating the on-shell bulk action \eqref{eq:bulkMax}  on the WdW patch, the boundary contributions are slightly subtle. One needs to specify the choice of ensemble at this point. We prefer to work in the grand canonical ensemble, which amounts to having a fixed chemical 
potential for the boundary CFT. For any other ensemble, one has to be careful about Maxwell boundary terms on different surfaces \cite{Chamblin:1999tk, Goto:2018iay}.  We postpone further discussion on this subtlety to Sec. \ref{sec:discussion-outlook}.

While the Gibbons-Hawking term is required to achieve a well-defined variational principle, to guarantee finite free energies in respective regions of spacetime one needs to add further counterterms \cite{Emparan:1999pm}. 
The explicit forms of these terms are given by\footnote{The same counterterm \eqref{I-CT-def} also appears for a flat boundary metric in the context of holographic renormalization \cite{deHaro:2000vlm}. 
} 
\begin{align}
I_{\rm{GH}} &= \pm \frac{1}{8 \pi G} \int d^{d+1} x \sqrt{|h|} K \label{I-GH-def}\\ 
I_{\rm{CT}} & =\mp  \frac{1}{8 \pi G} \int d^{d+1} x \sqrt{|h|} \frac{d}{L} \label{I-CT-def}.
\end{align}

Since the WdW patch possesses both spacelike and timelike boundary surfaces, we have to be careful about fixing the signs in front of \eqref{I-GH-def} and  \eqref{I-CT-def}. A timelike surface corresponds to the upper choice of sign while for a spacelike one, the lower sign is appropriate. For example, in our setup, the upper signs of both  \eqref{I-GH-def} and  \eqref{I-CT-def} are to be used for the cutoff at $r_c$, whereas the lower signs have to be used when dealing with the cutoff at $r_0$. 

The requirement of having boundary terms on the null boundaries can be avoided by simply choosing an affine parametrization of the null directions as we will do in what follows. However, for such boundaries, we do need to consider contributions to the action coming from joint points. These are the points where two null boundaries intersect or a null boundary intersects with a spacelike or timelike boundary.
The former case requires
\begin{align}
 I_{\rm j} &= \pm \frac{1}{8 \pi G} \int d^d x \sqrt{\sigma} \log{\frac{|v_1 \cdot v_2|}{2}},
\label{I-j-def}
\end{align}
where $v_1$ and $v_2$ are the null vectors of the two respective boundaries 
\begin{equation}\label{eq:lightlikevectors}
v_1=\alpha\left(-dt+\frac{dr}{f(r)} \right)\;\;\;v_2=\beta\left(dt+\frac{dr}{f(r)} \right)\,.
\end{equation}
Here, $\alpha$ and $\beta$ are parameters, which must be introduced due to the ambiguous nature of the normalization of lightlike vectors.  $\sigma$ is the induced metric on this surface.

The latter case, namely the intersection of a lightlike boundary with either a spacelike or timelike boundary takes on a similar form
\begin{align}
 I_{\rm j} &= \pm \frac{1}{8 \pi G} \int d^d x \sqrt{\sigma} \log{n^{\mu}v_{\mu}}\,,
\label{I-j2-def}
\end{align}
where $n^{\mu}$ refers to the unit normal of the timelike/spacelike surface and where $v_{\mu}$ is as given in \eqref{eq:lightlikevectors}.

In order to evaluate these contributions explicitly, it is convenient to introduce the tortoise coordinate $r^*$ as
\begin{equation}\label{eq:tortoisecoordinate}
r^*=-\int_{r}^{\infty}\frac{dr}{f(r)}\,,
\end{equation}
in terms of which, from Fig. \ref{fig:latetimecompgrowth}, we can read off the null boundaries of the WdW patch at late times as
\begin{align}\label{eq:lightlikeboundaries}
B_1: t&= t_R + r^*(r_c) - r^*(r)  \;\;\;\;\;\;\; B_3: t =-t_L +r^*(r_c) - r^*(r) \nonumber \\
B_2: t&= t_R -r^*(r_c) +r^*(r) \;\;\;\;\;\;\;B_4: t =-t_L -r^*(r_c) +r^*(r)  \,.
\end{align}

The action of the WdW patch only depends on the time $\tau = t_L + t_R$ and not on the individual boundary times, $t_L$ and $t_R$. This follows trivially from the boost symmetry. Therefore, to simplify our computation, without any loss of generality, we can consider a time-symmetric configuration, namely, $t_L  = t_R = \frac{t}{2}$.
One could, in principle, also choose time-shifted configurations, but this would not affect the late-time growth of complexity. 


\subsection{The bulk and the boundary contributions}
In order to evaluate the bulk contribution along with the Gibbons-Hawking and the counterterms, we split up half the WdW patch into three regions, $a$, $b$, and $c$, as depicted in Fig. \ref{fig:latetimecompgrowth}. The bulk and boundary terms of the full WdW patch will then be obtained by simply doubling the contributions. 

 Let us first work through the bulk contributions. We work on the full WdW patch (the union of the dark and the light green regions) before reducing to the intersection of WdW with the future interior. Evaluating the action, \eqref{eq:bulkMax}, on-shell in the three regions $a$, $b$, and $c$ yields 
 
 \begin{align}
 I_{\rm{bulk}}^a &= \frac{L^d V}{8 \pi G} \int^{r_0}_{r_+} dr \int_{0}^{B_1} dt\bigg( -\frac{(d+1)}{r^{d+2}} + Q^2(d-1) r^{d-2} \bigg) \nonumber \\
 &=  \frac{L^d V}{8 \pi G} \int^{r_0}_{r_+} dr \bigg( -\frac{(d+1)}{r^{d+2}} + Q^2(d-1) r^{d-2} \bigg) \bigg(\frac{\tau}{2} + r^*(r_c) - r^*(r) \bigg)\,,\label{eq:Iabulk}\\
 I_{\rm{bulk}}^b &= \frac{L^d V}{8 \pi G} \int^{r_+}_{r_c} dr \int_{B_2}^{B_1} dt \bigg( -\frac{(d+1)}{r^{d+2}} + Q^2(d-1) r^{d-2} \bigg)\nonumber\\
 &= \frac{L^d V}{8 \pi G} \int^{r_+}_{r_c} dr  \bigg( -\frac{(d+1)}{r^{d+2}} + Q^2(d-1) r^{d-2} \bigg)\bigg( r^*(r_c) - r^*(r) \bigg)\,,\label{eq:Ibbulk}\\
  I_{\rm{bulk}}^c &= \frac{L^d V}{8 \pi G} \int^{r_m}_{r_+} dr \int_{B_2}^{0}dt \bigg( -\frac{(d+1)}{r^{d+2}} + Q^2(d-1) r^{d-2} \bigg)\nonumber\\
  &= \frac{L^d V}{8 \pi G} \int^{r_m}_{r_+} dr \bigg( -\frac{(d+1)}{r^{d+2}} + Q^2(d-1) r^{d-2} \bigg) \bigg(-\frac{\tau}{2} + r^*(r_c) - r^*(r) \bigg)\label{eq:Icbulk} \,.
  \end{align}
  As stated above, for the full WdW patch these contributions have to be doubled. Note, that we are working in a late-time approximation, $r_m \approx r_+$, such that \eqref{eq:Icbulk} vanishes. Both the Gibbons-Hawking term \eqref{I-GH-def} and the counterterm \eqref{I-CT-def} do not contribute on the lightlike segments due to the affine parametrization we choose. However, both do appear at the spacelike cutoff surface located at $r_0$.
 Here we get
 \begin{align}\label{eq:WdWGHcounterterm1}
 I_{\rm{GH}} &=   - 2 \times \; \frac{V L^d  }{8 \pi G} \int_{0}^{\tau/2} dt \bigg( (d+1) \frac{1}{r_0^{d+1}} +  Q^2 r_0^{d-1} - \frac{1}{2}(d+1)m \bigg) \,,\\
 I_{\rm{CT}} &= 2 \times \;  \frac{V L^d  }{8 \pi G} \int_0^{\tau/2} dt \frac{d}{r_0^{d+1}} \sqrt{|f(r_0)|} \label{eq:WdWGHcounterterm2}\,.
 \end{align}  
The factors of $2$ appearing in front of \eqref{eq:WdWGHcounterterm1} and \eqref{eq:WdWGHcounterterm2} follow from the same logic of doubling the contributions.

 
 \subsection{Contributions from the joint points}
In Fig. \ref{fig:latetimecompgrowth}, as far as the late-time behavior is concerned, there are 
five joint point contributions, which we will evaluate using \eqref{I-j-def} and \eqref{I-j2-def}. 
In order to specify the location of these points it is useful to switch to the following coordinates 
\cite{Agon:2018zso}
\begin{equation}\label{eq:nullcoordinates}
u=-e^{-\frac{1}{2}f'(r_+)(r^*-t)}\,,\;\;\;v=-e^{-\frac{1}{2}f'(r_+)(r^*+t)}\,.
\end{equation}
In these coordinates the horizon is located at  $u v=0$ or equivalently $r^*(r_+)=-\infty$ on which three of the joint points are located. However, since both $r^*(r_+)$ and $\log (f(r_+))$ diverge, we must introduce $\epsilon_u$ and $\epsilon_v$, which can be interpreted as regularized locations of the horizon. The three lightlike joint points are then represented by
\begin{align}\label{eq:lightlikejunctions}
J_1&: (\epsilon_u,v_0)\,, \;\;\;\;\;\;\; J_2:(\epsilon_u,\epsilon_v)\,,  \;\;\;\;\;\;\;  J_3: (u_0,\epsilon_v) \,,
\end{align}
 where $v_0$ and $u_0$ designate the future interior null boundaries. We denote the  corresponding radial coordinates by $r_{u_0,\epsilon_v}$, $r_{\epsilon_u, v_0}$ and $r_{\epsilon_{u}, \epsilon_{v}}$ respectively. Using \eqref{I-j-def}, we can evaluate the contribution coming from these three joint points. Similarly, contributions from the other two joints located at the spacelike cutoff  $r_0$ can be evaluated using \eqref{I-j2-def}. The total contribution from all five joint points is given by
 
 \begin{equation}\label{eq:jointcontribution1}
 I_{\rm{j}}=\frac{V L^d}{8 \pi G}\left(\frac{\log \frac{\alpha\beta r_0^2}{L^2 |f(r_0)|}}{r_0^d}+\frac{\log \frac{\alpha \beta r_{\epsilon_{u}, \epsilon_{v}}^2}{L^2 |f(r_{\epsilon_{u}, \epsilon_{v}})|}  }{ r_{\epsilon_{u}, \epsilon_{v}}^d}-\frac{\log \frac{\alpha \beta r_{u_0 , \epsilon_{v}}^2}{L^2 |f(r_{u_0 , \epsilon_{v}})|}  }{ r_{u_0 , \epsilon_{v}}^d}-\frac{\log \frac{\alpha \beta r_{\epsilon_u,v_0}^2}{L^2 |f(r_{\epsilon_u, v_0})|}  }{ r_{\epsilon_u , v_0}^d} \right)\,,
 \end{equation}
where the first term corresponds to the two joint points at $r_0$ and the remaining three terms, to the lightlike joint points. We work in the approximation $r_{\epsilon_u, \epsilon_v}\approx r_+$,  $r_{u_0 , \epsilon_v}\approx r_+$ and $r_{\epsilon_u , v_0}\approx r_+$, such that \eqref{eq:jointcontribution1} simplifies to
\begin{equation}
\label{eqn:Ij}
I_{\rm{j}}=\frac{V L^d}{8 \pi G}\left( \frac{\log |f(r_{\epsilon_{u}, v_0})| +\log |f(r_{u_0,\epsilon_v})|-\log |f(r_{\epsilon_u,\epsilon_v})|}{r_+^d}   - \frac{\log \frac{\alpha \beta r_+^2}{L^2}}{r_+^d} -\frac{\log \frac{\alpha \beta r_0^2}{L^2 |f(r_0)|}}{r_0^d}                   \right)\,.
\end{equation}
Furthermore, in the limit, $u v \rightarrow 0$, $\log |f(r_{u,v})|$ appearing in \eqref{eqn:Ij} can be approximated as \cite{Agon:2018zso}
\begin{equation}\label{eq:nullapproximation}
\log |f(r_{u,v})|=\log |u v |+ c_0 + {\cal O}(uv) \,,
\end{equation}
where $c_0$ is an $u,v$ independent function. This further simplifies \eqref{eqn:Ij} to

\begin{equation}\label{eq:Ijunction}
 I_{\rm j}  = \frac{ V L^d}{8 \pi G} \left( \frac{\log | u_0 v_0| + c_0}{r_+^d} - \frac{\log |f(r_0)|}{r_0^d} - \frac{\log \frac{\alpha \beta r_+^2}{L^2} }{r_+^d} +  \frac{\log \frac{\alpha \beta r_0^2}{L^2} }{r_0^d} \right)\,.
\end{equation}
The ambiguity in \eqref{eq:Ijunction} due to the presence of the last two terms may in principle be removed by a further counterterm  \cite{Lehner_2016, Reynolds:2016rvl, Alishahiha:2019cib}. However, since we are only interested in the growth rate of complexity, we can ignore this issue since only the $\tau$-dependent first term of \eqref{eq:Ijunction} will contribute in this case.


\subsection{The late-time growth of complexity} Now that we have all the constituents appearing in \eqref{IWdW}, evaluated on shell, on the intersection of the WdW patch with the future interior, we can evaluate the late-time growth of the action by taking derivatives of \eqref{eq:Iabulk}, \eqref{eq:Ibbulk}, \eqref{eq:WdWGHcounterterm1}, \eqref{eq:WdWGHcounterterm2} and \eqref{eq:Ijunction} with respect to $\tau$
\begin{align}
\frac{d I_{\rm{bulk}}}{d \tau}&= \frac{L^{d} V}{8 \pi G} \left(\frac{1}{r_0^{d+1}}-\frac{1}{r_+^{d+1}}+Q^2(r_0^{d-1}-r_+^{d-1})\right)\label{eq:Cbulkdt}\,,\\
\frac{d I_{\rm{GH}}}{d \tau}&=\frac{L^{d} V}{8 \pi G} \left(\frac{m}{2}(1+d)-\frac{(1+d)}{r_0^{d+1}}-Q^2 r^{d-1}\right)\,,\\
\frac{d I_{\rm{CT}}}{d \tau}&=\frac{L^{d} V}{8 \pi G} \left(\frac{d \sqrt{-f(r_0)}}{r_0^{d+1}}\right)\,,\\
\frac{d I_{\rm j}}{d \tau}&=\frac{L^{d} V}{8 \pi G} \left(\frac{(1+d)m}{2}-d Q^2 r_+^{d-1}\right)\label{eq:Cjdt}\,.
\end{align}
On the other hand by making use of  \eqref{eq:CA}, one obtains the late-time growth of
 complexity as follows 
\begin{align}
\frac{d C}{d\tau} = \frac{L^{d} V}{8 \pi^2 G \hbar} \left\{ (d+1) m - \frac{d}{r_0^{d+1}} \bigg( 1- \sqrt{-f(r_0)} \bigg) - \frac{1}{r_+^{d+1}} - Q^2 (d+1) r_+^{d-1} \right\}\,,
\label{eq:latetimecompgr}
\end{align}
which could be further simplified, using  $f(r_+)=0$, to find
\begin{align}
\frac{d C}{d\tau} = \frac{L^{d} Vd}{8 \pi^2 G \hbar} \left\{  \frac{1}{r_+^{d+1}}  +\frac{1}{r_0^{d+1}} \bigg( \sqrt{-f(r_0)}\;-1 \bigg) \right\}\,,
\label{eq:latetimecompgr}
\end{align}
which has the same form as that of the neutral case (see \cite{Akhavan:2018wla}) and the only charge 
dependence comes from the blacking factor $f(r_0)$. It is evident that it reduces to 
that of the neutral case in the zero charge limit. On the other hand it is also clear that for 
 $r_0\rightarrow r_-$ it gives the standard expression for the late-time growth of complexity of a 
 charged black brane \cite{Carmi:2017jqz, Cai:2016xho},
 \begin{align}
\frac{d C}{d\tau} = \frac{L^{d} Vd}{8 \pi^2 G \hbar} \left(  \frac{1}{r_+^{d+1}} 
 -\frac{1}{r_-^{d+1}}  \right)=\frac{L^{d} Vd}{8 \pi^2 G \hbar}Q^2\left(r_-^{d-1}-r_+^{d-1}\right) \,.
\end{align} 
 
It is believed that the late-time behavior of complexity should be expressed in terms of 
conserved charges such as energy. Therefore there must be a relation between $r_0$ and
the UV cutoff $r_c$, so that the above expression for the late-time growth of complexity can be written 
entirely in terms of the conserved charges defined at the boundary. To find such a 
relation it is natural to use the Lloyd's bound. To do so, in the next section we revisit Lloyd's bound
for charged black branes (black holes). 
 

\section{LLOYD'S BOUND AND BEHIND THE HORIZON CUTOFFS } 
\label{sec:Lloyd}
Lloyd's bound  constitutes an upper bound on the growth rate of the quantum complexity of any physical system \cite{Lloyd_2000}. Explicitly it is given as
\begin{equation}\label{eq:generalLLyodsbound}
\frac{dC}{dt} \leq \frac{2 {\cal E}}{\pi \hbar} \, ,
\end{equation}
$\cal E$ being the energy of the system.

In \cite{Brown:2015bva, Brown:2015lvg} it was shown that the late-time growth of the WdW action of a charge neutral AdS black hole satisfies the relation 
  \begin{equation}
  \label{bound-Susskind}
 \frac{d {\cal I}_{\rm{WdW}}}{d\tau} = 2 {\cal E}\, = 2 M\,,
 \end{equation}
 which, in conjunction with \eqref{eq:CA}, can be read as a statement on the saturation of Lloyd's bound \eqref{eq:generalLLyodsbound}. $M$ here is the ADM mass of the black hole in AdS which in the context of the AdS/CFT correspondence can be identified with the energy of the boundary CFT.\footnote{Although we are calling it  ``Lloyd's bound" for historical reasons, more appropriately, we should think of this as a bound on holographic complexity given in terms of conserved quantities at the holographic boundary. In presence of a UV cutoff, this bound will, therefore, be sensitive to the cutoff.}. 
 
 They further proposed the generalization of Lloyd's bound for a charged AdS black hole as 
 \begin{equation}\label{eq:generalLLyodsbound-charged}
\left(\frac{dC}{dt}\right)_{\rm{Lloyd}} \leq \frac{2}{\pi \hbar} \left[\left({\cal E} - \mu {\cal Q}\right) - \left({\cal E}  - \mu {\cal Q}\right)_{\rm{gs}} \right]
\, ,
\end{equation}
 where the subscript ${\rm gs}$ denotes the ground state.  Computing the growth explicitly and using the CA proposal, they concluded that while this bound is saturated for a small charged black hole, it is violated for finite-sized charged black holes.  
 
 However, the authors of \cite{Cai:2016xho} showed explicitly that the claim of  \cite{Brown:2015bva, Brown:2015lvg} was actually incorrect and even small charged black holes disobey the bound \eqref{eq:generalLLyodsbound-charged}. This observation was quite intriguing because \cite{Brown:2015bva, Brown:2015lvg} attributed the saturation and violation of the bound, respectively, by small and finite sized charged black holes to the fact that the bound could only be saturated in the supersymmetric limit and for UV complete holographic theories. This further fueled the possibility that the bound proposed in \cite{Brown:2015bva, Brown:2015lvg} was actually inappropriate.
 
 Due to the fast scrambling nature of black holes,\footnote{Black holes may be considered the fastest scramblers, see \cite{Sekino:2008he}.}, it is natural to expect that black holes should 
 saturate the appropriately defined bound, irrespective of its size. In \cite{Cai:2016xho} a new version of the bound on charged black holes was proposed, namely
 
  \begin{equation}\label{eq:generalLLyodsbound-charged-Cai}
\frac{dC}{dt}  \leq \frac{1}{\pi \hbar} \left[\left({\cal E} - \mu_+ {\cal Q}\right) - 
\left({\cal E}  - \mu_- {\cal Q}\right)\right]
\, ,
\end{equation}
where $\cal E$ is the total energy and ${\cal Q}$ is the ADM charge of the system given in 
\eqref{ADMQ}.  $\mu_\pm$ denote, ``formally", the chemical potentials associated with the two 
horizons, $r_+$ and $r_-$ respectively. However, one might complain that this statement is a bit 
vague since the thermodynamics of the inner horizon is not a strictly well-understood concept, and 
rightly so. The way we should understand $\mu_-$ here is through the replacement $r_+ 
\rightarrow r_-$ in the expression for the CFT chemical potential $\mu_+$ which is 
thermodynamically well defined. However, it is worth mentioning that, although $\mu_-$ cannot be 
interpreted as a chemical potential from the perspective of the boundary CFT, it can be 
reexpressed in terms of other well-defined boundary quantities using the expressions of ADM mass 
and charge in terms of $r_+$ and $r_-$ \cite{Cai:2016xho}.  

While \cite{Cai:2016xho} explicitly established the saturation of the proposed Lloyd's bound 
\eqref{eq:generalLLyodsbound-charged-Cai} for charged (and also rotating and Gauss-Bonnet) 
AdS black holes, they also showed how this proposal reduces to \eqref{bound-Susskind} in the 
uncharged limit. This limit is very interesting and follows from the fact that in the neutral limit, where 
$Q \rightarrow 0$ and $r_- \rightarrow \infty$ simultaneously, $\mu_+ Q \rightarrow 0$ and $\mu_- Q 
\rightarrow 2 M$ while the energies $\cal E$ cancel between the two terms.

Although \eqref{eq:generalLLyodsbound-charged-Cai} conforms with the saturation of Lloyd's 
bound for charged AdS black holes of any size, along with other categories of black holes, this is 
slightly unnatural. This can be understood from the limiting argument to the neutral case 
mentioned above. The contribution in this limit arises solely from the ``-" side, namely from the term 
which modifies the original proposal for Lloyd's bound because of the existence of the inner 
horizon. It is legitimate to expect that all such contributions coming from the ``-" side will add up to 
zero in the neutral limit. 

Moreover, due to this unnatural limiting behavior, it was argued that Eq.\eqref{eq:generalLLyodsbound-charged-Cai}
 is unable to accommodate the right expression for Lloyd's bound when a finite cutoff is applied to the theory \cite{Hashemi:2019xeq}.\footnote{The authors 
 of \cite{Hashemi:2019xeq} used the equation \eqref{eq:generalLLyodsbound-charged}
 for Lloyd's bound.}.

Based on the above observations we would like  to propose a new bound for the late-time growth of complexity as follows:

\be
\label{eq:ourlloydsbound}
\left(\frac{dC}{dt}\right)_{\rm{bound}} =  \frac{1}{\pi \hbar }\left[\left(2 {\cal E}_+ - \mu_+ {\cal Q}\right) - \left(2 {\cal E}_- - \mu_- {\cal Q}\right) \right],
\ee

with $\pm$ denoting the quantities associated with the outer and inner horizons as before. 
It is then clear that in the zero-charge limit, the contributions coming from the ``-" side will add up to 
zero and we are left with the contribution of the ``+" side. 
Of course for most cases in which ${\cal E}_+={\cal E}_-$ both proposals 
\eqref{eq:generalLLyodsbound-charged-Cai} and \eqref{eq:ourlloydsbound} result in the same 
expression for Lloyd's bound. The difference shows up when ${\cal E}_+\neq {\cal E}_-$, which may happen when we have a finite cutoff. Indeed, we will also see that the expression
for Lloyd's bound as given in \eqref{eq:ourlloydsbound} allows to consider the 
theory in the presence of finite cutoffs. As we mentioned before following \eqref{eq:generalLLyodsbound-charged-Cai}, it is worth stressing once again that ${\mu}_-$ cannot be directly interpreted as a chemical potential. However, it is possible to express ${\mu}_-$ in terms of well-defined conserved charges of the boundary CFT \cite{Cai:2016xho}.

To proceed we note that for the case in which there is no cutoff, the ADM energy can be directly computed from the on shell gravitational action as \cite{Chamblin:1999tk} 
\be
\label{energy0}
{\cal E}_0= \frac{L^{d}V  d}{16\pi G}m,
\ee 
which is the same for both the inner and the outer horizon. Now if we consider the cutoff at $r=r_c$, the physical energy gets corrected by \cite{Hartman:2018tkw}
\be
\label{energy-cutoff}
\frac{L^{d}V  d}{16\pi G}m\rightarrow \frac{L^{d} V}{8 \pi G} \frac{d}{r_c^{d+1}} 
\left(1 - \sqrt{|f(r_c)|}\right).
\ee
This can be derived in two steps, first by computing the bulk energy enclosed by the cutoff surface at $r = r_c$ and then using the holographic dictionary for the cutoff AdS/ $TT$-deformed CFT correspondence to compute the boundary energy\footnote{See \cite{Hartman:2018tkw} for the details of this computation.}
\be
\label{eqn:energy-cutoff+}
{\cal E}_+ = \frac{L^{d} V}{8 \pi G} \frac{d}{r_c^{d+1}} 
\left(1 - \sqrt{|f(r_c)|}\right).
\ee
While the presence of the cutoff will only modify the energy contribution in the outside ``$+$'' region, the contribution coming from the ``$-$'' region will remain unaffected,
\be
\label{eqn:energy-cutoff-}
{\cal E}_- =\frac{L^{d}V  d}{16\pi G}m\,\, .
\ee
This is supported by the zero charge limit in which, as we mentioned,
one would naturally expect that the contribution of the ``-" side should vanish in this limit.

We also need the contributions to Lloyd's bound coming from the chemical potentials. 
Following our conventions in \eqref{eq:RNmetric}, \eqref{eq:RNgauge}, and in \eqref{ADMQ}, the chemical potentials for the outer and the inner horizons are  given by
\be
\label{chempotpm}
\mu {\cal Q}|_+=
\frac{L^{d} Vd}{8 \pi G}Q^2 r_+^{d-1} ,\;\;\;\;\;\;\;\;\;\mu {\cal Q}|_-= 
\frac{L^{d} Vd}{8 \pi G}Q^2 r_-^{d-1} \,.
\ee
Now we are in a position to combine  \eqref{eqn:energy-cutoff+}, \eqref{eqn:energy-cutoff-} and \eqref{chempotpm} to evaluate the bound given in \eqref{eq:ourlloydsbound}
\begin{align}
\label{LBdef1}
\left(\frac{dC}{d\tau}\right)_{\rm{bound}} &= \frac{L^{d} Vd}{8 \pi^2 \hbar G} \left[2 
\frac{1}{r_c^{d+1}} \left(1 - \sqrt{|f(r_c)|}\right)
 -Q^2 r_+^{d-1} \right] - \frac{L^{d} Vd}{8 \pi^2 \hbar G} \left[m 
-Q^2 r_-^{d-1} \right]\nonumber\\
&=\frac{L^{d} Vd}{8 \pi^2 \hbar  G} \left[ \frac{ 2}{r_c^{d+1}} \left(1 - \sqrt{|f(r_c)|}\right)  - {m} 
 +Q^2 (r_{-}^{d-1} -r_{+}^{d-1}) \right] \,.
\end{align}
From this expression, it is clear that in the limit $Q\rightarrow 0$, the full 
contribution coming from the ``-'' side, namely $\left(2 {\cal E}  - \mu {\cal Q}\right)_{-}$, vanishes 
identically leaving only the contribution coming from the canonical energy ${\cal E}_+$ at the 
boundary. This leads to the expected result
\begin{equation}
\left(\frac{dC}{d\tau}\right)_{\rm{bound}} =\frac{L^{d} Vd}{8 \pi^2 \hbar  G} \left[ \frac{ 2}{r_c^{d+1}} 
\left(1 - \sqrt{|f(r_c)|}\right)  \right]=2{\cal E}_+\,,
\end{equation}

It is worth stressing again that  both proposals for a suitable Lloyd's bound for a charged black 
brane, the one considered in \cite{Cai:2016xho} and the other we proposed, namely, 
\eqref{eq:ourlloydsbound}, reduce to the uncharged limit, \eqref{bound-Susskind}. However, the limits are achieved in crucially different manners. Contrary to our case discussed above, in the neutral limit of \cite{Cai:2016xho}, ``-" quantities contribute.

\subsection{Behind the horizon cutoff}
Apart from our proposal being physically more reasonable, it will turn out that in the presence of cutoffs, our proposal is the apt one. Before justifying this in the following section, let us use our 
proposal to find the relation between the two cutoffs.

Equating  \eqref{LBdef1} to the late-time growth of complexity \eqref{eq:latetimecompgr}, one 
obtains a relation between $r_c$ and $r_0$ as follows: 
\be \label{eq:cutoffrelation}
  \frac{ 2}{r_c^{d+1}} \left(1 - \sqrt{|f(r_c)|}\right)  - {m} 
 =   \frac{1}{r_-^{d+1}}+\frac{1}{r_0^{d+1}} \bigg( \sqrt{-f(r_0)}\;-1 \bigg) \,.
\ee
To write this equation we have used the fact that
\be
  \frac{1}{r_+^{d+1}}  -\frac{1}{r_-^{d+1}} =Q^2\left(r_-^{d-1}-r_+^{d-1}\right) \,.
\ee

In the next section, we will use a different approach to arrive at the result \eqref{eq:cutoffrelation}, which implicitly also verifies \eqref{eq:ourlloydsbound}.


\section{FACTORIZATION OF THE PARTITION FUNCTION}
\label{sec:factorization}
In this section we will investigate the possibility of relating the UV and the behind the horizon cutoff from a completely different perspective. This study follows directly from the 
construction of interior operators in terms of those describing  the exterior regions  
proposed in \cite{Papadodimas:2012aq}. Actually, based on this fact one can argue 
that the partition function of the operators describing  
the interior of an eternal black hole is proportional to the product 
of partition functions of operators describing the left and right exteriors of the black hole \cite{Akhavan:2019vtt}. At leading order this connection may be reduced to a relation between the on-shell actions evaluated on the inside and outside of the black hole. 

Using this approach for neutral, eternal black holes
one may find an expression  for  the cutoff behind the horizon which is the same as 
that obtained in the  context of holographic complexity \cite{Akhavan:2019vtt}.
The aim of this section is to extend this study to charged black branes.

\begin{figure}[h!]
\centering
\begin{tikzpicture}
[scale=0.65] 
	\node (I)    at ( 4,0)   {};
\node (II)   at (-4,0)   {};
\node (IIIa)  at (0, 2.5) {};
\node (IV)   at (0,-2.5) {};
\node (IIIb) at (0, 5) {};
\node (V) at (-4,7.9) {};
\node (VI) at (4,7.9) {};

\node   at (0, 2.5) {II};
\node   at (0,-2.5) {IV};
\node at (-1.8,0)   {III};
 	 \node at (1.8,0)   {I};
 	 \node at (1.8,8)   {VI};
 	 \node at (-1.8,8)   {V};

\path  
  (VI) +(90:4)  coordinate[]  (VItop)
       +(-90:4) coordinate[] (VIbot)
       +(0:-4)   coordinate                  (VIleft);
       \draw 
      (VItop) --
          node[midway, below, sloped] {}
      (VIleft) ;
      \draw (VIleft) -- 
           node[midway, below, sloped] {}
      (VIbot);
\draw    (VIbot) [decorate,decoration=snake]    --
          node[midway, above, sloped] {}
          node[midway, below left]    {}    
      (VItop) ;
\path  
  (V) +(90:4)  coordinate[]  (Vtop)
       +(-90:4) coordinate[] (Vbot)
       +(0:4)   coordinate                  (Vright);
       \draw 
      (Vtop) --
          node[midway, below, sloped] {}
      (Vright) -- 
          node[midway, below, sloped] {}
      (Vbot) ;
\draw   (Vbot)  [decorate,decoration=snake]    --
          node[midway, above, sloped] {}
          node[midway, below left]    {}    
      (Vtop) ;

\path  
  (II) +(90:4)  coordinate[]  (IItop)
       +(-90:4) coordinate[] (IIbot)
       +(0:4)   coordinate                  (IIright);
       \draw 
      (IItop) --
          node[midway, below, sloped] {}
      (IIright) -- 
          node[midway, below, sloped] {}
      (IIbot) --
          node[midway, above, sloped] {}
          node[midway, below left]    {}    
      (IItop) -- cycle;

\path 
   (I) +(90:4)  coordinate (Itop)
       +(-90:4) coordinate (Ibot)
       +(180:4) coordinate (Ileft)
       ;
\draw  (Ileft)-- (Itop)-- (Ibot) -- (Ileft) -- cycle;


      \draw[blue,decorate, decoration={segment length=1cm, pre=lineto, pre length=0.6cm, post = lineto, post length=0.1cm}] (IItop)  to[bend right=13]  (Itop);
      \node[blue] at (0,3.1) {$r_0$};
      \draw[red,decorate, decoration={segment length=1cm, pre=lineto, pre length=0.6cm, post = lineto, post length=0.1cm}] (Itop)  to[bend right=15]  (Ibot);
      \node at (2.85,0)[red] {$r_c$};
       \node at (-2.85,0)[red] {$r_c$};
 	
 	 \draw[red,decorate, decoration={segment length=1cm, pre=lineto, pre length=0.6cm, post = lineto, post length=0.1cm}] (IItop)  to[bend left=15]  (IIbot);
 	 
 	 \node at (5.5,0) {$r=0$};
 	 \node at (-5.5,0) {$r=0$};
 	  \node at (5.5,8) {$r\rightarrow \infty$};
 	 \node at (-5.5,8) {$r\rightarrow \infty$};
 	 \node at (-2.2,-3) {$r_+$};
 	 \node at (2.2,-3) {$r_+$};
 	 \node at (2.2,9.3) {$r_-$};
 	 \node at (-2.2,9.3) {$r_-$};
 	 
\end{tikzpicture}

\caption{Penrose diagram of the eternal, charged black brane. The radial cutoff in region $I$ lies at $r=r_c$ and induces a cutoff behind the outer horizon at $r=r_0$. }
\label{fig:PenroseEternalBB}
\end{figure}
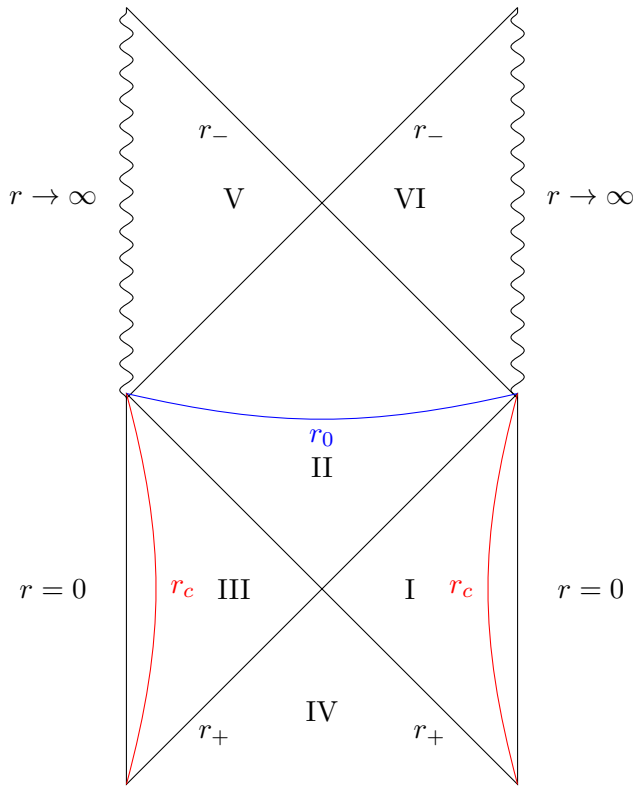

To proceed, we note that  in regions $\mathrm I$ and $\mathrm{III}$ of the Penrose diagram depicted in Fig. \ref{fig:PenroseEternalBB}, AdS/CFT provides us with a map allowing us to write down nonlocal CFT operators 
 playing the role of the local bulk fields in these regions
\begin{align}
\label{exp-I-III}
\Phi^{{\mathrm I}}_{\mathrm{CFT}} &= \int_{\omega>0} \frac{d\omega \, d^d k}{\left(2 \pi\right)^{d+1}} \left[{\cal O}_{\omega, {\vec k}} \, f_{\omega, {\vec k}} \left(t,{\vec x}, r \right) + \hbox{H.c.} \right], \nonumber \\
\Phi^{\mathrm{III}}_{\mathrm{CFT}} &= \int_{\omega>0} \frac{d\omega \, d^d k}{\left(2 \pi\right)^{d+1}} \left[{\tilde{\cal O}}_{\omega, {\vec k}} \, f_{\omega, {\vec k}} \left(t,{\vec x}, r \right) + \hbox{H.c.} \right],
\end{align}
where ${\cal O}_{\omega, {\vec k}}$ and ${\tilde{\cal O}}_{\omega, {\vec k}}$ are Fourier transforms of generalized free fields in the CFT. These are special CFT operators whose $n$-point correlators factorize into $2$-point correlators at  leading order in the large $N$ expansion \cite{ElShowk:2011ag}.  The mode functions $ f_{\omega, {\vec k}}$ are the solutions of the bulk equations of motion in these two regions subject to normalizability conditions at the respective boundaries.

On the other hand in the interior regions, such as $\mathrm{II}$,  representation of the local bulk field needs both sets of operators \cite{Papadodimas:2012aq}
\begin{equation}
\label{exp-II}
\Phi^{\mathrm {II}}_{\mathrm{CFT}} = \int_{\omega>0} \frac{d\omega \, d^d k}{\left(2 \pi\right)^{d+1}} \left[{\cal O}_{\omega, {\vec k}} \, g_{\omega, {\vec k}} \left(t,{\vec x}, r \right) + {\tilde{\cal O}}_{\omega, {\vec k}} \, {\tilde g}_{\omega, {\vec k}} \left(t,{\vec x}, r \right) + \mathrm{H.c.} \right],
\end{equation}
where  $g^i_{\omega, {\vec k}} \left(t,{\vec x}, r \right)$ and ${\tilde g}^i_{\omega, {\vec k}} \left(t,{\vec x}, r \right)$ are bulk mode functions in respective regions. However, for the obvious reason that these regions cannot access the AdS boundaries, one cannot impose any boundary conditions on these solutions, and \eqref{exp-II} follows naturally from the smoothness of the horizon \cite{Papadodimas:2012aq} or equivalently, the entanglement structure of the dual CFT state \cite{VanRaamsdonk:2009ar}. It is worth mentioning that in the expansions \eqref{exp-I-III} and \eqref{exp-II}, the bulk radial coordinate $r$, plays the role of a nonlocality parameter in the dual CFT.

Following \cite{Akhavan:2019vtt}, let us define the {\it restricted}
 partition function in which  the integration is 
 taken over the fields  associated with 
 regions ${\mathrm I}$,  ${\mathrm {II}}$, or $\mathrm{III}$ of the corresponding eternal black hole

\begin{equation}
{\cal Z}^{({\mathrm \alpha})} \propto \int {\cal D}\Phi^{\alpha} e^{-i S^{(\alpha)}\left[\Phi^\alpha \right]},
\end{equation} 
where $\alpha = \{{\mathrm I},  {\mathrm {II}}, \mathrm{III}\}$.  
We can rewrite the path integral using the mode expansions in the respective regions which yield
\begin{align}
{\cal Z}^{({\mathrm {I}})} & \propto  \int {\cal D} {\cal O}_{\omega, \vec{k}} \; {\cal D} {\cal O}_{-\omega, -\vec{k}} \; e^{-i S^{(\mathrm {I})}[{\cal O}]}\\
{\cal Z}^{({\mathrm {III}})} & \propto  \int {\cal D} {\tilde{\cal O}}_{\omega, {\vec k}} \; {\cal D} {\tilde{\cal O}}_{-\omega, {-\vec k}} \; e^{-i S^{(\mathrm {III})}[{\cal \tilde O}]}\\
{\cal Z}^{({\mathrm {II}})} & \propto  \int {\cal D} {\cal O}_{\omega, \vec{k}} \;  {\cal D} {\tilde{\cal O}}_{\omega, {\vec k}} \; {\cal D} {\cal O}_{-\omega, -\vec{k}} \; {\cal D} {\tilde{\cal O}}_{-\omega, {-\vec k}} \; e^{-i S^{(\mathrm {II})}[{\cal O}, {\cal {\tilde O}}]}\,.
\label{eq:partII}
\end{align}
In general the restricted partition function in region II, \eqref{eq:partII}, does not factorize into the contributions coming from the modes $\cal O$ and $\cal{\tilde O}$ : $S^{(\mathrm {II})}[{\cal O}, {\cal {\tilde O}}] \ne S^{(\mathrm {I})}[{\cal O}] + S^{(\mathrm {III})}[{\cal {\tilde O}}]$. However, we know that for generalized free fields, mixed correlators factorize at leading order of  the $1\over N$ expansion \cite{ElShowk:2011ag},
\begin{align}
\langle {\cal O}_1  {\cal O}_2 \cdots {\cal O}_n  {\cal \tilde O}_1  {\cal \tilde O}_2 \cdots {\cal \tilde O}_m \rangle &= \frac{1}{{\cal Z}^{({\mathrm {II}})}} \frac{d^{n+m} {\cal Z}^{({\mathrm {II}})}}{d J^n d{\tilde J}^m} \bigg\rvert_{J={\tilde J} = 0} \nonumber\\
&=  \frac{1}{{\cal Z}^{({\mathrm {I}})}} \frac{d^{n} {\cal Z}^{({\mathrm {I}})}}{d J^n} \bigg\rvert_{J= 0} + \frac{1}{{\cal Z}^{({\mathrm {III}})}} \frac{d^{m} {\cal Z}^{({\mathrm {III}})}}{d{\tilde J}^m} \bigg\rvert_{{\tilde J} = 0}  + O \left(1\over N\right) \nonumber\\
& = \langle {\cal O}_1  {\cal O}_2 \cdots {\cal O}_n\rangle \langle {\cal \tilde O}_1  {\cal \tilde O}_2 \cdots {\cal \tilde O}_m \rangle + O \left(1\over N\right).
\end{align}
Therefore in the large $N$ limit, one obtains a simple relation between the restricted partition functions at leading order of the $1\over N$ expansion
\begin{equation}
\label{restrictedZrel}
{\cal Z}^{({\mathrm {II}})} \propto {\cal Z}^{({\mathrm I})}{\cal Z}^{(\mathrm{III })}\,.
\end{equation}

Intuitively, the above equation indicates that  in order to study region $\mathrm {II}$,
 one needs twice the number of  modes as those in region ${\mathrm {I}}$.  It is worth noting that since the operator \eqref{exp-II} is a nonlocal 
 operator in the dual field theory whose nonlocality parameter is given by the AdS 
 redial coordinate, imposing any restriction on the nonlocality parameter
  (such as setting a UV cutoff) would restrict the range of the spacetime accessible 
  to those fields defined behind the horizon.

Therefore, assuming that spacetime is cut off at the radial distance $r_c$, immediately and 
automatically implies that there should also be a second cutoff in region $\mathrm{II}$, the region 
behind the outer horizon. This provides a justification for the existence of the interior cutoff ``$r_0$", we introduced in the context of late-time growth of the WdW action before. Furthermore, one can 
fix the proportionality constant in \eqref{restrictedZrel} and write down a relation involving the 
on-shell actions, in the respective regions of the Penrose diagram as 
\begin{align}
\label{restrictedSrel}
e^{i(S_{\mathrm{cut-off}}^{(\mathrm{II})} - S_0^{(\mathrm{II})})} = e^{2i(S_{\mathrm{cut-off}}^{(\mathrm{I})} - S_0^{(\mathrm{I})})},
\end{align}
with $S_0^{(\mathrm{i})}$ denoting the on-shell action evaluated without a cutoff. 
 In view of the fact that the original relation \eqref{restrictedZrel} was derived for generalized free fields, one might wonder at this point about how one can identify the restricted actions with the gravitational actions in respective regions.\footnote{We would like to thank the anonymous referee for bringing up this subtlety.}. An intuitive justification for doing this comes from the fact that at leading order of $1 \over N$, the classical effective action is indeed given by the Einstein-Hilbert-Maxwell action, \eqref{eq:bulkMax}. The fluctuation of this action around the classical geometry given by \eqref{eq:RNmetric} and \eqref{eq:RNgauge} gives rise to the expectation value of the graviton field which can be treated as a generalized free field of the dual CFT. However, we should mention that this is only an intuitive argument and we do not have a concrete proof for the same. Rather we will consider this as a proposition motivated by \cite{Akhavan:2019vtt} where the relation between the restricted gravitational actions led to a relation between two cutoffs in the case of an uncharged black brane which was found to be perfectly consistent with the well-accepted Lloyd's bound in the uncharged case. It is legitimate to expect that the same should also work for our charged black brane.

Now we would like to use the relation \eqref{restrictedSrel} to determine the relation between the two 
cutoffs: the UV cutoff, $r_c$ and the cutoff behind the horizon, $r_0$. We will again assume the 
interior cutoff $r_0$ to lie between the inner and the outer horizons. Furthermore, here too we will compute 
the on-shell actions in the grand canonical ensemble. Also, as is standard procedure, in order to ensure finite free energies in all regions we are required to use both the Gibbons-Hawking terms and the counterterms which 
have the forms given in \eqref{I-GH-def} and \eqref{I-CT-def} respectively. With this, let us now 
write down the on-shell actions in different regions explicitly.

\subsubsection*{Region $\mathrm I$: Outside the outer horizon}
First we calculate the on-shell action in the region outside the outer horizon. This entails a radial integration from $r_c$ to $r_+$. The individual components of the on-shell action are given by
\bea
S_{\mathrm{bulk}}^{\rm{(I)}}(r_c)
&=&\frac{L^{d}V_d \tau}{8\pi G}
\left(\frac{1}{r_+^{d+1}}+Q^2r_+^{d-1}\right)-\frac{L^{d}V_d \tau}{8\pi G}
\left(\frac{1}{r_c^{d+1}}+Q^2r_c^{d-1}\right),\cr&&\cr
&=&\frac{L^{d}V_d \tau}{8\pi G}m-\frac{L^{d}V_d \tau}{8\pi G}
\left(\frac{1}{r_c^{d+1}}+Q^2r_c^{d-1}\right),\cr&&\cr
S_{\rm{GH}}^{\rm{(I)}}(r_c)&=&\frac{L^dV_d\tau}{8\pi G}\left(\frac{d+1}{r_c^{d+1}}+Q^2 r_c^{d-1}-
\frac{d+1}{2}m\right)\,,\cr&&\cr
S_{\rm{CT}}^{\rm{(I)}}(r_c)&=&- \frac{L^d V_d\tau}{8\pi G}
\frac{d}{r_c^{d+1}}\sqrt{f(r_c)}\,.
\eea
Here, we have also introduced $\tau$ as a cutoff in time direction. By summing up the individual contributions we arrive at the total on-shell action in region $\rm I$ 
\bea
\label{S-II}
S^{\rm{(I)}}(r_c) &=& S_{\mathrm{bulk}}^{\rm{(I)}}(r_c) + S_{\rm{GH}}^{\rm{(I)}}(r_c) + S_{\rm{CT}}^{\rm{(I)}}(r_c)\nonumber \\
&=&\frac{L^{d}V_d \tau}{16\pi G}(1-d)m+\frac{L^{d}V_d \tau}{8\pi G}
\frac{d}{r_c^{d+1}}\left(1- \sqrt{f(r_c)}\right)\,.
\eea
We now normalize this expression with respect to the no cutoff case. Hence, we subtract from \eqref{S-II}, the asymptotic boundary limit, $r_c=\epsilon\rightarrow 0$, namely,
\bea
S^{\rm{(I)}}(\epsilon)
=\frac{L^{d}V_d \tau}{16\pi G}m\,.
\eea
This yields
\bea
\label{eq:deltas1}
\Delta S^{\rm{(I)}} =S^{\rm{(I)}}(r_c) - S^{\rm{(I)}}(\epsilon)
&=&-\frac{L^{d}V_d \tau d}{16\pi G}m+\frac{L^{d}V_d \tau}{8\pi G}
\frac{d}{r_c^{d+1}}\left(1- \sqrt{f(r_c)}\right)\, .
\eea

\subsubsection*{Regions $\mathrm{II}$: Between the two horizons}
We now move to region $\rm{II}$, which in principle runs from $r_{+}$ to $r_{-}$. However, we are assuming the existence of a cutoff situated between $r_-$ and $r_+$.  Hence, we perform a radial integration from $r_+$ to $r_0$. 
First note that without a cutoff, the bulk action in region $\rm{II}$ amounts to
\bea
S_{\mathrm{bulk}}^{\rm{(II)}}
&=&\frac{L^{d}V_d \tau}{8\pi G}
\left(\frac{1}{r_-^{d+1}}+Q^2r_-^{d-1}\right)-\frac{L^{d}V_d \tau}{8\pi G}
\left(\frac{1}{r_+^{d+1}}+Q^2r_+^{d-1}\right)=0.
\eea
However, if we set a cutoff at  $r_0<r_-$ this changes to
\bea
S_{\mathrm{bulk}}^{\rm{(II)}}(r_0)
&=&\frac{L^{d}V_d \tau}{8\pi G}
\left(\frac{1}{r_0^{d+1}}+Q^2r_0^{d-1}\right)-\frac{L^{d}V_d \tau}{8\pi G}m,
\eea
which, as expected, clearly vanishes as we set $r_0=r_-$. 
Noting again the spacelike nature of the cutoff surface $r_0$ in region $\rm{II}$, 
 the contributions of the boundary terms  \eqref{I-GH-def}, \eqref{I-CT-def} are given by
\bea
S_{\rm GH}^{\rm{(II)}}(r_0)&=&-\frac{L^dV_d\tau}{8\pi G}\left(\frac{d+1}{r_0^{d+1}}+Q^2 r_0^{d-1}-
\frac{d+1}{2}m\right)\,,\cr&&\cr
S_{\rm{CT}}^{\rm{(II)}}(r_0)&=& \frac{L^d V_d\tau}{8\pi G}
\frac{d}{r_0^{d+1}}\sqrt{-f(r_0)}\,.
\eea

Hence, the full on-shell action in the interior for $r_0<r_-$ is given by
\bea
S^{\rm{(II)}}(r_0) &=& S_{\mathrm{bulk}}^{\rm{(II)}}(r_0) + S_{\rm{GH}}^{\rm{(II)}}(r_0) + S_{\rm{CT}}^{\rm{(II)}}(r_0)\nonumber \\
&=&\frac{L^{d}V_d \tau}{16\pi G}(d-1)m
+ \frac{L^d V_d\tau}{8\pi G}
\frac{d}{r_0^{d+1}}\left(\sqrt{-f(r_0)}-1\right)\,.
\eea
 
Just as we did for region $\mathrm I$, we want to normalize the cutoff partition function by subtracting the asymptotic contribution, which for this case amounts to $r_0=r_-$. Setting $r_0=r_-$ makes the counter term and also the bulk contribution vanish, the Gibbons-Hawking term remains nonzero yielding the full on-shell action in this region without cutoff as
\be
S^{\rm{(II)}}(r_-) =-\frac{L^dV_d\tau}{8\pi G}\left(\frac{d}{r_-^{d+1}}-
\frac{d-1}{2}m\right)\,.
\ee

One can then evaluate the difference $\Delta S^{\rm{(II)}}$ as
\bea
\label{eq:deltas2}
\Delta S^{\rm{(II)}} &=& S^{\rm{(II)}}(r_0) - S^{\rm{(II)}}(r_-) \nonumber \\
&=& \frac{L^dV_d\tau d}{8\pi G}\left\{\frac{1}{r_-^{d+1}}+
\frac{1}{r_0^{d+1}}\left(\sqrt{-f(r_0)}-1\right)\right\}.
\eea

We can now simply use  \eqref{eq:deltas1} and \eqref{eq:deltas2} in \eqref{restrictedSrel} to find a relation between the two cutoffs: $r_0$ and $r_c$ 
\begin{align}
\label{relation-partition}
\frac{2}{r_c^{d+1}} \left( 1 - \sqrt{|f(r_c)|} \right)-m =\frac{1}{r_{-}^{d+1}}+\frac{1}{r_{0}^{d+1}}\left( \sqrt{-f(r_0)} - 1\right)\,.
\end{align}

This relation is identical to the one obtained by demanding the saturation of Lloyd's bound in \eqref{eq:cutoffrelation}. This matching is very much reminiscent of the chargeless case already noted in  \cite{Akhavan:2019vtt}. The charged scenario, in presence of the inner horizon will provide a new interpretation of this result.

\section{TOWARDS A HOLOGRAPHIC REALIZATION OF STRONG COSMIC CENSORSHIP}
\label{sec:SCC}
Let us come back to the issue of the location of the cutoff $r_0$.  As stated previously, there are, in principle, two options in placing this cutoff. One may either put it between the inner and the outer horizons, or it may also be assumed to lie behind the inner horizon. In all our computations presented above, we chose the former option. Although this was only a choice initially, we will now argue that this choice leads to a self-consistent physical interpretation of our result.

It is worth mentioning here that in \cite{Alishahiha:2019cib}, the interior cutoff
was assumed to lie behind the 
inner horizon.\footnote{We note however, that the main conclusion of that paper was not based 
on this assumption.}. This leads to the late-time growth of complexity being independent of this cutoff. 
Of course this was also consistent with Lloyd's bound \eqref{eq:generalLLyodsbound-charged-Cai} upon which this assumption was made.
In this paper we have argued that this expression for Lloyd's bound exhibits certain unnatural
features and therefore needs to be modified. 

Moreover the assumption made in \cite{Alishahiha:2019cib} results in inconsistencies if the growth has to approach and eventually saturate a bound at late times. This is because, whatever the correct bound is, it should depend on the physical, thermodynamical quantities of the boundary CFT and as we already saw, these quantities are explicitly dependent on the UV cutoff, $r_c$. As a result, there will be a mismatch of scales if we aim to construct an equation describing a bound, $\cal B$, on the growth of complexity, namely,

\begin{equation}\label{eq:generalbound}
\frac{dC}{dt} \leq {\cal B} \, ,
\end{equation}

\noindent where the left-hand side of \eqref{eq:generalbound} will now be independent of any cutoff and the right-hand side will be dependent on $r_c$. One could argue that this might be the case if saturation is never met. However, it is quite unreasonable to expect such a situation for an AdS black hole due to the ``fast scrambler" argument mentioned before. 

Another interesting aspect of the setup considered in \cite{Alishahiha:2019cib} is that it adds to an apparent ambiguity. While it was assumed that the late-time growth of complexity is independent of the cutoff behind the horizon, in order to achieve the expected complexity growth for the near horizon AdS${_2}$ region, one does need to consider the counterterms coming from the behind the horizon cutoff. Therefore, it turns out, in the setup of \cite{Alishahiha:2019cib}, this interior cutoff is very much essential but it is not clear how this should get fixed explicitly in terms of the UV cutoff, contrary to our expectations.

Having the interior cutoff between the two horizons solves all the aforementioned problems in a consistent way. With the interior cutoff placed between the inner and the outer horizons, the left-hand side of \eqref{eq:generalbound} depends on $r_0$ and the right-hand side, on $r_c$, thus providing the relation between the cutoffs. Moreover, we obtained exactly the same relation from the factorization of the partition function. As a consequence of having a well-defined bulk effective field theory, the factorization is expected to be obeyed at least when the cutoff is sufficiently close to the boundary of AdS.

The aforementioned arguments in favour of having the interior cutoff in between the inner and the outer horizons give us a hint about bulk reconstruction in AdS/CFT. The exact matching of the relation between the cutoffs makes it clear that complexity, as a probe, cannot penetrate the inner horizon of the black brane or black hole if it is to be consistent with the factorization of the Hilbert space at large $N$. It therefore indicates emergence of a holographic censorship in bulk reconstruction behind the inner horizon. One might naturally identify this as a version of strong cosmic censorship arising from holography.

\section{DISCUSSION AND OUTLOOK}
\label{sec:discussion-outlook}

In this work, we have revisited holographic complexity for charged black branes in the
presence of a finite cutoff. We have seen that a UV finite cutoff enforces a cutoff behind the outer horizon, with  an expression determined by the UV cutoff. This was shown in two ways. 

First, by assuming that Lloyd's bound is saturated for a charged black hole in the presence of a cutoff, we related conserved charges of this system to the late-time growth of complexity. Here, interestingly, the charges are only sensitive to the UV cutoff, whereas the late-time behavior of holographic complexity seems blind to $r_c$. Assuming an agreement to hold at late times forces a relation between the two cutoffs.

Secondly, we saw that the same relation may also be obtained using the leading-order factorization of the partition function in the $1\over N$ expansion. Following Papadodimas-Raju's
construction of interior operators in terms of exterior operators, implies that behind the horizon, twice the number of modes are required.

A crucial point in order to make the overall setup consistent is to use the correct expression for Lloyd's bound in terms of conserved charges such as the electric charge at the boundary. Although in the literature there are several proposals for Lloyd's bound, they suffer from certain pathologies.
Our proposal \eqref{eq:ourlloydsbound},  in the neutral limit, reduces to the Schwarzschild case in a more natural way with the contributions arising solely from the outer horizon. Furthermore and perhaps more importantly, our proposed relation between the two cutoffs \eqref{eq:cutoffrelation} as derived using our proposal for Lloyd's bound,\eqref{eq:ourlloydsbound}, guarantees an exact match with the relation obtained from the Papadodimas-Raju construction.

Our proposal for the expression of Lloyd's bound  \eqref{eq:ourlloydsbound}, can be further generalized to systems with more physical conserved charges. For instance, in the case of a charged rotating system, it can be readily generalized to
\be \label{eq:generalllodysbound}
\frac{dC}{dt}\leq \frac{1}{\pi \hbar}(2{\cal E}-\mu Q-\Omega J)_+-(2{\cal E}-\mu Q-
\Omega J)_-\,,
\ee
where $J$ and $ Q$ are angular momentum and charge, and $\Omega$ and $\mu$ 
are their corresponding potentials.

In this paper we have only considered the case of an electric black brane, 
although we could have also considered dyonic black holes, in which the system carries both 
electric and magnetic charges. While in this case the final expressions become more 
involved, essentially the physical conclusions remain unchanged.

Moreover, in our computations, we have made a specific choice of ensemble. We are working in a grand canonical ensemble, in which the chemical potential $\mu$ is considered fixed. However, it is of course interesting to examine different choices of ensemble, specifically the canonical ensemble.  This requires the addition of a boundary term to the action \eqref{eq:bulkMax}
\begin{align}
\label{boundary-Maxwell}
S_{\rm{M, b}} = \frac{\gamma}{8 \pi G} \int d^{d+1} x \sqrt{|h|} n_{\mu} F^{\mu \nu} A_{\nu} \,.
\end{align}
This generalizes our calculations to a larger choice of ensembles designated by values of $\gamma$, which is in the interval $[0,1]$.  $\gamma = 1$ corresponds to the canonical ensemble where the total charge $Q$ is held fixed, while $\gamma = 0$ corresponds to the grand canonical ensemble where instead the chemical potential $\mu$ of the system is fixed and hence reduces to the approach outlined in this paper. Choices in between correspond to mixed ensembles. Holographic complexity of charged black holes in the presence of this boundary term has been studied in \cite{Brown:2018bms, Goto:2018iay}.

The boundary term \eqref{boundary-Maxwell} would alter both the on-shell action computed on the WdW patch and also the computations of partition functions taking into account the effective field theory of the interior.  Accordingly, the relations \eqref{eq:cutoffrelation} and \eqref{relation-partition} between $r_c$ and $r_0$ should be generalized to an arbitrary choice of ensemble.  The natural question would then be if the relations obtained from the two approaches should  agree for  general $\gamma$ or if this should only work for a specific choice of ensemble. We are investigating this issue and we hope to come back with a precise answer in a future publication. It will also be interesting to understand the near horizon, near extremal limit of the construction with arbitrary $\gamma$, as the presence of a boundary term of the form \eqref{boundary-Maxwell} was shown to be essential in this limit \cite{Alishahiha:2019cib, Brown:2018bms}.

Assuming that the leading order factorization of the partition function works for arbitrary
$\gamma$, following Sec. 4, the relation \eqref{relation-partition} generalizes to
\be\label{relation-partitiona}
\frac{2}{r_c^{d+1}} \left( 1 - \sqrt{|f(r_c)|} \right)-m =\frac{1}{r_{-}^{d+1}}+\frac{1}{r_{0}^{d+1}}\left( \sqrt{-f(r_0)} - 1\right)-\gamma Q^2 \left( r_0^{d-1}-r_{-}^{d-1}+2r_c^{d-1}\right).
\ee
It is interesting to see if this expression is consistent with the complexity computations for general ensembles.

Another aspect which we would like to investigate in the future is the ``factorization puzzle", namely, the apparent tension of the exact factorization of the boundary Hilbert space and the loss thereof due to bulk wormhole structures \cite{Harlow:2015lma, Harlow:2018tqv, Saad:2019lba, Saad:2021rcu, Verlinde:2021jwu}. Following the connection between the bound on the late-time growth of complexity and the factorization of the partition function that we developed in this work, it is of course highly interesting to see in how far the presence of bulk wormholes is captured by the growth of complexity. Particularly, connecting to the discussion above, it will be worth investigating if the ensemble dependence plays any crucial role here.


\acknowledgments
We would like to thank Johanna Erdmenger, Ren\'e Meyer and Ali Naseh for several useful discussions. We would like to thank Johanna Erdmenger for comments on the manuscript. The work of S.B. is supported by the Alexander von Humboldt postdoctoral fellowship. The work of J.K.K. was supported in part by the Heising-Simons Foundation, the Simons Foundation, and National Science Foundation Grant No. NSF PHY-1748958.

\bibliographystyle{JHEP}
\bibliography{reference}

\end{document}